\documentclass[12pt]{article}
\usepackage{pdproc}
\usepackage{epsfig}

\def\a{\alpha}
\def\b{\beta}

\def\d{\delta}
\def\e{\epsilon}

\def\g{\gamma}
\def\h{\eta}

\def\j{\psi}
\def\k{\kappa}
\def\l{\lambda}
\def\m{\mu}
\def\n{\nu}

\def\r{\rho}
\def\s{\sigma}
\def\t{\tau}

\def\z{\zeta}
\def\D{\Delta}
\def\F{\Phi}

\def\L{\Lambda}

\def\Q{\Theta}

\def\ve{\varepsilon}

\def\dg{\dagger}                                     
\def\wt#1{\widetilde{#1}}                    
\def\VEV#1{\left\langle #1\right\rangle}        
\def\beq{\begin{equation}}
\def\eeq{\end{equation}}

\def\bea{\begin{eqnarray}}
\def\eea{\end{eqnarray}}
\def\NO{\nonumber}
\def\Bar#1{\overline{#1}}

\def\pl#1#2#3{Phys.~Lett.~{\bf B {#1}} ({#2}) #3}
\def\np#1#2#3{Nucl.~Phys.~{\bf B {#1}} ({#2}) #3}
\def\prl#1#2#3{Phys.~Rev.~Lett.~{\bf #1} ({#2}) #3}
\def\pr#1#2#3{Phys.~Rev.~{\bf D {#1}} ({#2}) #3}

\def\ap#1#2#3{Ann.~of Phys.~{\bf {#1}} ({#2}) #3}
\def\prep#1#2#3{Phys.~Rep.~{\bf {#1}C} ({#2}) #3}

\begin{document}
\pagestyle{empty}
\begin{flushright}
{DESY 01-186}\\
\end{flushright}
\vspace*{1cm}
\begin{center}
{\bf NEUTRINO MASSES IN GUTS \\ AND THE BARYON ASYMMETRY} \\
\vspace*{1cm} 
{\bf W. Buchm\"uller} \\
\vspace{0.3cm}
Deutsches Elektronen-Synchrotron DESY, 22603 Hamburg, Germany\\
\vspace*{2cm}  
{\bf ABSTRACT} \\ \end{center}
\vspace*{10mm}
\noindent
We discuss the implications of large neutrino mixings for grand unified 
theories based on the seesaw mechanism. In SU(5) GUTs large mixings 
can be accomodated by means of $U(1)_F$ flavour symmetries. In these
models the heavy Majorana neutrinos are essentially decoupled from 
low energy neutrino physics. On the contrary in SO(10) GUTs
large neutrino mixings severely constrain the mass spectrum
of the heavy Majorana neutrinos. This leads to predictions for a variety
of observables in neutrino physics as well as for the 
cosmological baryon asymmetry.

\vspace*{5cm} 

\begin{center}
{\it Presented at the 8th Adriatic Meeting\\ 
`Particle Physics in the New Millenium',
Dubrovnik, September 2001,\\ 
and at the XXV International School of Theoretical Physics \\
`Particles and Astrophysics - Standard Models and Beyond', 
Ustro\'n, September 2001}
\end{center}

\vfill\eject

\setcounter{page}{2}
\pagestyle{plain}

\renewcommand{\thefootnote}{\alph{footnote}}
  
\normalsize\baselineskip=15pt

\section{Status of neutrino mixing}

Recent results from the Sudbury Neutrino Observatory\cite{sno} and from
the Super-Kamiokande experiment\cite{sk01} provide further evidence for 
neutrino oscillations as the solution of the solar neutrino
problem. Neutrino oscillations can also account for the atmospheric neutrino 
anomaly\cite{sk98,k2k}. It is remarkable that a consistent picture can be 
obtained with just three neutrinos, $\nu_e$, $\nu_\mu$ and $\nu_\tau$,  
undergoing `nearest neighbour' oscillations,  
$\nu_e \leftrightarrow \nu_\mu$ and $\nu_\mu \leftrightarrow \nu_\tau$.

For massive neutrinos a mixing matrix $U$ appears in the leptonic charged 
current,
\beq
{\cal L}_{CC} = -{g\over \sqrt{2}}\sum_{\a,i} 
\overline{e}_{\a} \g^\m (1 - \g_5) U_{\a,i} \n_i\ W^-_\m \ + \ldots\;,
\eeq
where $e_\a$ and $\n_i$ are mass eigenstates. In the case of three neutrinos,
one for each generation, $U$ is a unitary matrix.

The experimental results on the  $\nu_e$ deficit in the solar neutrino flux 
favour the LMA or LOW solutions\cite{bax98} of  the MSW conversion with 
large mixing angle. A large mixing also fits the atmospheric neutrino oscillations.
As a result, the leptonic mixing matrix
$U_{\alpha i}$ appears to be very different from the familiar CKM quark 
mixing matrix $V_{\alpha i}$. The emerging pattern is rather 
simple\cite{ft01},
\beq\label{umix}
U = \left(\begin{array}{ccc}
    \ast  & \ast  & \diamond \\
    \ast  & \ast  & \ast \\
    \ast  & \ast  & \ast 
    \end{array}\right) \;.
\eeq
Here the `$\ast$' denotes matrix elements whose value is consistent with the 
range $0.5 \ldots 0.8$, whereas for the matrix element 
`$\diamond$' only an upper bound exits, $|U_{e3}| < 0.16$. The neutrino 
masses may be hierarchical or quasi-degenerate. Note, however, that a 
possible hierarchy has to be much weaker than the known mass hierarchy of 
quarks and charged leptons.

Several interesting phenomenological schemes have been suggested, such as
`bi-maximal' or `democratic' mixing, which describe the pattern (\ref{umix}) 
rather well\cite{rev00}. Is is unclear, however, how these schemes 
are related to a more fundamental theory. 
We shall therefore focus on the question how large neutrino mixings 
can be obtained in a grand unified theory based on the gauge groups SU(5) 
or SO(10). In both cases we shall rely on 
the seesaw mechanism which naturally explains the smallness of light Majorana 
neutrino masses $m_\n$ by the largeness of right-handed neutrino masses 
$M$\cite{seesaw},  
\beq
m_{\nu}\simeq - m_D{1\over M}m_D^T \;, \label{seesaw}
\eeq
where $m_D$ is the Dirac neutrino mass matrix. In unified theories 
$m_D$ is related to the quark and charged lepton mass
matrices. Since they have a large hierarchy, the almost non-hierarchical
structure of the leptonic mixing matrix is very surprising and requires
some explanation. In the following we shall discuss two qualitatively
different examples based on the GUT groups SU(5) and SO(10), respectively,
which illustrate present attempts to solve the puzzle of the large
neutrino mixings.

\section{Models with SU(5)}

In the simplest GUT based on the gauge group SU(5)\cite{gg74}
quarks and leptons are grouped into the multiplets
${\bf 10}=(q_L,u_R^c,e_R^c)$, ${\bf 5^*}=(d_R^c,l_L)$ and ${\bf 1}= \n_R$.
Hence, unlike the gauge fields, quarks and leptons are not unified in a single
irreducible representation. In particular, the right-handed neutrinos are
gauge singlets and can therefore have Majorana masses not generated by
spontaneous symmetry breaking. In addition one has three Yukawa interactions, 
which couple the fermions to the Higgs fields $H_1({\bf 5})$ and 
$H_2({\bf 5^*})$, 
\beq  
{\cal L} = h_{uij} {\bf 10}_i {\bf 10}_j H_1({\bf 5})
          +h_{dij} {\bf 5^*}_i {\bf 10}_j H_2({\bf 5^*}) 
          +h_{\n ij} {\bf 5^*}_i {\bf 1}_j H_1({\bf 5})
          + M_{ij} {\bf 1}_i {\bf 1}_j \;.  
\eeq
The mass matrices of up-quarks, down-quarks, charged leptons and the Dirac
neutrino mass matrix are given by
$m_u = h_u v_1$, $m_d = h_d v_2$, $m_e = m_d$ and $m_D = h_\n v_1$, 
respectively, with $v_1 = \VEV H_1$ and $v_2 = \VEV H_2$. 
The Majorana masses $M$ are independent of the
Higgs mechanism and can therefore be much larger than the electroweak scale
$v$.

An attractive framework to explain the observed mass hierarchies of quarks
and charged leptons is the Froggatt-Nielsen mechanism\cite{fn79} based
on a spontaneously broken U(1)$_F$ generation symmetry. 
The Yukawa couplings are assumed to
arise from non-renormalizable interactions after a gauge singlet field $\F$ 
acquires a vacuum expectation value,
\beq
h_{ij} = g_{ij} \left({\VEV\F\over \L}\right)^{Q_i + Q_j}\;.
\eeq
Here $g_{ij}$ are couplings ${\cal O}(1)$ and $Q_i$ are the U(1)$_F$ 
charges of the
various fermions, with $Q_{\F}=-1$. The interaction scale $\L$ is
usually chosen to be very large, $\L > \L_{GUT}$. 
\begin{table}[b]

\begin{center}
\begin{tabular}{c|ccccccccc}\hline \hline
$\j_i$  & $ \bf 10_3 $ & $ \bf 10_2 $ & $ \bf 10_1 $ & $ \bf 5^*_3 $ & 
$ \bf 5^*_2 $ & $\bf 5^*_1 $ & $ \bf 1_3 $ & $ \bf 1_2 $ & $ \bf 1_1 $ 
\\\hline
$Q_i$  & 0 & 1 & 2 & $a$ & $a$ & $a+1$ & b & $c$ & $d$ \\ \hline\hline
\end{tabular}
\end{center}
\caption{\it Lopsided U(1)$_F$ charges of SU(5) multiplets.
From \protect\cite{by99}.}
\end{table}

The symmetry group SU(5)$\times$U(1)$_F$ has been considered by a number of 
authors. Particularly interesting is the case with a `lopsided' family 
structure where the chiral U(1)$_F$ charges are different for the 
$\bf 5^*$-plets and the $\bf 10$-plets of the same 
family\cite{sy98,ilr98,bw87}. Note, that such lopsided charge assignments are not
consistent with the embedding into a higher-dimensional gauge group, like
SO(10)$\times$U(1)$_F$ or E$_6\times$U(1)$_F$.
An example of phenomenologically allowed lopsided charges $Q_i$ is given in table~1.

This charge assignement determines the structure of the Yukawa matrices,
e.g.,
\beq\label{yuk}
h_e = h_d
  \ \sim\  \left(\begin{array}{ccc}
    \e^3 & \e^2 & \e^2 \\
    \e^2 &\;  \e \;   & \e   \\
    \e   & 1    & 1
    \end{array}\right) \;,
\eeq
where the parameter $\e = \VEV\F/\L$ controls the flavour mixing, and
coefficients ${\cal O}(1)$ are unknown.
The corresponding mass hierarchies for up-quarks, down-quarks and charged
leptons  are
\bea
\qquad\quad 
m_t : m_c : m_u  &\simeq & 1 : \e^2 : \e^4\;, \\
m_b : m_s : m_d  &=& m_\t : m_\m : m_e \simeq 1 : \e : \e^3\;.
\eea
The differences between the observed down-quark mass hierarchy and the
charged lepton mass hierarchy can be accounted for by introducing 
additional Higgs fields\cite{gj79}. From a fit to the running quark and
lepton masses at the GUT scale one obtains for the flavour mixing parameter 
$\e \simeq 0.06$.

The light neutrino mass matrix is obtained from the seesaw formula,
\beq\label{neuma}
m_{\n} = -m_D{1\over M}m_D^T \ \sim\ \e^{2a}\ \left(\begin{array}{ccc}
    \e^2  & \e  & \e \\
    \e  & \; 1 \; & 1 \\
    \e  &  1  & 1 
    \end{array}\right)\;.
\eeq
Note, that the structure of this matrix is determined by the 
U(1)$_F$ charges of the $\bf 5^*$-plets only. It is independent of the 
U(1)$_F$ charges of the right-handed neutrinos. 

Since all elements of the 2-3 submatrix of (\ref{neuma}) are ${\cal O}(1)$,
one naturally obtains a large $\n_\m -\n_\t$ mixing angle 
$\Q_{23}$\cite{sy98,ilr98}. At first sight 
one may expect that $\Q_{12} = {\cal O}(\e)$, which would correspond to
the SMA solution of the MSW conversion. However, one can also have
a large mixing angle $\Q_{12}$ if the determinent of the 2-3 submatrix
of $m_\n$ is ${\cal O}(\e)$\cite{vis98}. Choosing the coefficients
${\cal O}(1)$ randomly, in the spirit of `flavour anarchy'\cite{hmw00}, 
the SMA and the LMA solutions are about equally probable for 
$\e \simeq 0.1$\cite{sy00}. 
The corresponding neutrino masses are consistent with
$m_2 \sim 5\times 10^{-3}$~eV and $m_3 \sim 5\times 10^{-2}$~eV. 
We conclude that the neutrino mass matrix (\ref{neuma}) 
naturally yields a large angle $\Q_{23}$, with $\Q_{12}$ large or small. 
In order to have maximal mixings the coefficients ${\cal O}(1)$ have to
obey special relations.  

The model can also explain the cosmological baryon asymmetry via
leptogenesis\cite{fy86} for an appropriate choice of the parameters in 
table~1\cite{by99}. The mass of the heaviest Majorana neutrino is 
\beq
M_3\ \sim\ \e^{2(a+b)}\ {v_1^2\over\overline{m}_\n}\ 
\sim\ \e^{2(a+b)}\ 10^{15}\ \mbox{GeV}\;,
\eeq
where $\overline{m}_\n = \sqrt{m_2m_3} \sim 10^{-2}$~eV.
The special choice $a=b=0$, $c=1$, $d=2$ yields the
scenario of \cite{bp96} where $B-L$ is broken at the GUT scale. 

For the CP asymmetry in the decays of the heavy neutrinos $N_1$,
\beq
\ve_1 =
{\Gamma(N_1\rightarrow l \, H_2)-\Gamma(N_1\rightarrow l^c \, H_2^c)\over
\Gamma(N_1\rightarrow l \, H_2)+\Gamma(N_1\rightarrow l^c \, H_2^c)}\;,
\eeq    
one has in the case $M_1 < M_{2,3}$,
\beq
\ve_1 \simeq - {3\over 16\pi} {M_1\over (h_\n^\dg h_\n)_{11}}
 \mbox{Im}\left(h_\n^\dg h_\n {1\over M} h_\n^T h_\n^*\right)_{11} 
\sim  {3\over 16\pi} \e^{2(a+d)}\;.
\eeq
Successful baryogenesis requires $a+d=2$. With $\e \sim 0.1$ the
corresponding CP asymmetry is $\ve_1 \sim 10^{-6}$.
The baryogenesis temperature is then
$T_B \sim M_1 \sim \e^4 M_3 \sim 10^{10}$~GeV.
The effective neutrino mass which controls the out-of-equilibrium condition
of the decaying heavy Majorana neutrino is given by
$\widetilde{m}_1 = (m_D^\dagger m_D)_{11}/M_1 \sim 10^{-2}$~eV.

Thermal leptogenesis leads to the baryon asymmetry\cite{bp00} 
\beq\label{basym}
Y_B = {n_B-n_{\Bar{B}}\over s} = \k c_S {\ve_1\over g_*}\;,
\eeq
where $n_B$ and $s$ are baryon number and entropy densities, respectively;
$g_* \sim 100$ is the number of degrees of freedom in the plasma of the
early universe and $c_S = {\cal O}(1)$ is the conversion factor from
lepton asymmetry to baryon asymmetry due to sphaleron processes. 
Washout processes are accounted for by $\k <1$, which can be computed
by solving the full Boltzmann equations \cite{lut92,plu97}.
The resulting baryon asymmetry then reads
\beq
Y_B \sim \k\ 10^{-8}\;.
\eeq
With $\k \sim 0.1\ldots 0.01$ this is indeed the correct order of magnitude
in accord with observation, $Y_B \simeq (0.6 - 1)\times 10^{-10}$.

The magnitude for the generated baryon asymmetry depends crucially on the 
parameters $\ve_1$, $\widetilde{m}_1$ and $M_1$. In the models with
SU(5)$\times$U(1)$_F$ symmetry low energy neutrino physics is essentially
decoupled from the heavy Majorana neutrinos and does not constrain the
value of $M_1$. Hence, successful baryogenesis is  
consistent with the SU(5)$\times$U(1)$_F$ symmetry, but it cannot be
considered a generic prediction.
This is different in unified theories with larger gauge groups.

\section{Models with SO(10)}

The simplest grand unified theory which unifies one generation of
quarks and leptons including the right-handed neutrino in a single 
irreducible representation is based on the gauge group SO(10)\cite{gfm75}.
The quark and lepton mass matrices are obtained from the couplings of the 
fermion  multiplet ${\bf 16}=(q_L,u_R^c,e_R^c,d_R^c,l_L,\n_R)$ to the Higgs 
multiplets $H_1({\bf 10})$, $H_2({\bf 10})$ and $\Phi({\bf 126})$,
\beq  
{\cal L} = h_{uij} {\bf 16}_i {\bf 16}_j H_1({\bf 10})
          +h_{dij} {\bf 16}_i {\bf 16}_j H_2({\bf 10})
          +h_{Nij} {\bf 16}_i {\bf 16}_j \Phi({\bf 126})\;.
\eeq
Here we have assumed that the two Higgs doublets of the standard model
are contained in the two\footnote{
Note, that this is unavoidable in models with SO(10) breaking by orbifold 
compactification\cite{abc01}.} ten-plets $H_1$ and $H_1$, respectively.
This yields the quark mass matrices $m_u = h_u v_1$, $m_d = h_d v_2$, 
with $v_1 = \VEV H_1$ and $v_2 = \VEV H_2$, and the 
lepton mass matrices
\beq\label{muni}
m_D = m_u  \;, \quad m_e = m_d \;.
\eeq
Contrary to SU(5) GUTs, the Dirac neutrino and the up-quark mass matrices
are now related. Note, that all matrices are symmetric. The Majorana mass 
matrix $M = h_N \VEV \Phi$ is also generated by spontaneous symmetry breaking 
and a priori independent of $m_u$ and $m_d$. 

With $m_D = m_u$ the seesaw mass relation becomes
\beq
m_\n \simeq - m_u {1\over M} m_u^T \;.
\eeq
The large neutrino mixings now appear very puzzling, since the quark mass 
matrices are hierarchical and the quark mixings are small. It turns out,
however, that because of the known properties of the up-quark mass matrix
this puzzle can be resolved provided the heavy neutrino masses also obey 
a specific hierarchy. This then leads to predictions for a number of 
observables in neutrino physics including the cosmological baryon asymmetry.
In the following we shall describe these implications of large neutrino
mixings in SO(10) GUTs following ref.\cite{bw01}. 
The role of the heavy neutrino mass hierarchy for the light 
neutrino mixings has previously been discussed in different 
contexts\cite{smi93}.

From the phenomenology of weak decays we know that the quark matrices have
approximately the form\cite{fx00,rr01},
\beq\label{qmass}
m_{u,d} \propto \left(\begin{array}{ccc}
    0  & \e^3 e^{i\phi}  & 0 \\
    \e^3 e^{i\phi}  & \; \r\e^2 \; & \h\e^2 \\
    0  &  \h\e^2  & e^{i\j} 
    \end{array}\right) \;.
\eeq
Here $\e \ll 1$ is the parameter which determines the flavour
mixing, and 
\beq
\r = |\r| e^{i\a}\;, \quad \h = |\h| e^{i\b}\;,
\eeq
are complex parameters ${\cal O}(1)$.
We have chosen a `hierarchical' basis, where off-diagonal
matrix elements are small compared to the product of the corresponding
eigenvalues, $|m_{ij}|^2 \leq {\cal O}(|m_i m_j|)$. In contrast to the 
usual assumption of hermitian mass matrices\cite{fx00,rr01}, 
SO(10) invariance dictates the matrices to be symmetric.
All parameters may take different values
for up$-$ and down$-$quarks. Typical choices for $\e$ are $\e_u = 0.07$, 
$\e_d = 0.21$\cite{rr01}. The agreement with data can be improved by
adding in the 1-3 element a term ${\cal O}(\e^4)$\cite{beg00,rrx01}
which, however, is not important for the following analysis. Data also 
imply one product of phases to be `maximal', i.e., 
$\Delta = \phi_u-\a_u - \phi_d + \a_d \simeq \pi/2$.

We do not know the structure of the Majorana mass matrix $M = h_N \VEV \Phi$.
However, in models with family symmetries it should be similar to the quark mass 
matrices, i.e., the structure should be independent of the Higgs field.
In this case, one expects 
\beq\label{Mtext}
M = \left(\begin{array}{ccc}
    0  & M_{12}  & 0 \\
    M_{12}  & M_{22} \; & M_{23} \\
    0  &  M_{23}  & M_{33} 
    \end{array}\right) \;,
\eeq
with $M_{12} \ll M_{22} \sim M_{23} \ll M_{33}$. $M$ is diagonalized by
a unitary matrix, $U^{(N)\dg} M U^{(N)*} = \mbox{diag}(M_1,M_2,M_3)$.
Using the seesaw formula
one can now evaluate the light neutrino mass matrix. Since the choice of
the Majorana matrix $m_N$ fixes a basis for the right-handed neutrinos the
allowed phase redefinitions of the Dirac mass matrix $m_D$ are restricted.
In eq.~(\ref{qmass}) the phases of all matrix elements have therefore been 
kept. 

The $\n_\m$-$\n_\t$ mixing angle is known to be large. This leads us to
require $m_{\n_{i,j}}={\cal O}(1)$ for $i,j =2,3$. It is remarkable that this
determines the hierarchy of the heavy Majorana mass matrix to be\footnote{We
also note that this result is independent of the zeroes in the mass matrix
(\ref{qmass}) if its 1-3 element is smaller than $\e^3$, as required by
data.}
\beq
M_{12} : M_{22} : M_{33} = \e^5 : \e^4 : 1\;.
\eeq  
With $M_{33} \simeq M_3$, $M_{22} = \s \e^4 M_3$, 
$M_{23} = \z \e^4 M_3 \sim M_{22}$ and
$M_{12} = \e^5 M_3$, one obtains for masses and mixings to order 
${\cal O}(\e^4)$,
\beq\label{Mhier}
M_1 \simeq - {\e^6\over \s}  M_3\;, \quad M_2 \simeq \s \e^4 M_3\;,
\eeq
\beq\label{Mmix}
U^{(N)}_{12} = - U^{(N)}_{21} = {\e\over \s}\;,\quad
U^{(N)}_{23} = {\cal O}(\e^4)\;, \quad
U^{(N)}_{13} = 0\;.
\eeq
Note, that $\s$ can always be chosen real whereas $\z$ is in general complex.
This yields for the light neutrino mass matrix
\beq\label{nmass1}
 m_{\n}  = - \left(\begin{array}{ccc}
    0  & \e e^{2i\phi}  & 0\\
    \e e^{2i\phi}   & -\s e^{2i\phi} + 2\r e^{i\phi}   & \h e^{i\phi}  \\
    0  & \h e^{i\phi}   & \; e^{2i\j}
    \end{array}\right)\ {v_1^2 \over M_3}  \;.
\eeq
The complex parameter
$\z$ does not enter because of the hierarchy.
The matrix (\ref{nmass1}) has the same structure as the mass matrix 
(\ref{neuma}) in the $SU(5)\times U(1)_F$ model, except for additional 
texture zeroes.
Since, as required, all elements of the 2-3 submatrix are ${\cal O}(1)$, 
the mixing angle $\Q_{23}$ is naturally large. A large mixing 
angle $\Q_{12}$ can again occur in case of a small determinant of the
2-3 submatrix,
\beq\label{det}
(-\s + 2\r e^{-i\phi}) e^{2i\j} - \h^2 \equiv \d e^{2i\g} 
 = {\cal O}(\e)\;.
\eeq
Such a condition can be fullfilled without fine tuning if $\s, \r, 
\h = {\cal O}(1)$. It implies relations between the moduli as well as the 
phases of $\r$ and $\h$. In the special case of a somewhat smaller mass of the second
heavy neutrino, i.e., $|\s| < |\r|$, the condition (\ref{det}) becomes
\beq\label{special}
\j - \b \simeq {1\over 2} (\phi - \a)\;, \quad |\h|^2 \simeq 2 |\r|\;.
\eeq

The mass matrix $m_\n$ can again be diagonalized by a unitary matrix, 
$U^{(\n)\dg} m_{\n} U^{(\n)*}  = \mbox{diag}(m_1,m_2,m_3)$.
A straightforward calculation yields ($s_{ij} = \sin{\Q_{ij}}$,
$c_{ij} = \cos{\Q_{ij}}$, $\xi=\e/(1+|\h|^2)$),
\beq\label{nmix}
U^{(\n)}  = \left(\begin{array}{ccc}
 c_{12} e^{i(\phi-\b+\j-\g)}  & s_{12} e^{i(\phi-\b+\j-\g)}  & 
             \xi s_{23} e^{i(\phi-\b+\j)}\\
 - c_{23}s_{12} e^{i(\phi+\b-\j+\g)} & c_{23}c_{12} e^{i(\phi+\b-\j+\g)}  & 
              s_{23} e^{i(\phi+\b-\j)}  \\
  s_{23}s_{12} e^{i(\g+\j)} & -s_{23}c_{12} e^{i(\g+\j)}   & c_{23} e^{i\j}
    \end{array}\right) \;,
\eeq
with the mixing angles,
\beq
\tan{2\Q_{23}} \simeq {2|\h|\over 1-|\h|^2}\;, \quad
\tan{2\Q_{12}} \simeq 2 \sqrt{1+|\h|^2} {\e \over \d}\;.
\eeq
Note, that the 1-3 element of the mixing matrix is small, 
$U^{(\n)}_{13} = {\cal O}(\e)$. 
The masses of the light neutrinos are
\bea
m_1 \simeq  - \tan^2{\Q_{12}}\ m_2\;,\quad
m_2 \simeq {\e \over (1+|\h|^2)^{3/2}} \cot{ \Q_{12}}\ m_3\;,\quad
m_3 \simeq (1+|\h|^2)\ {v_1^2\over M_3}\;.
\eea 
This corresponds to the weak hierarchy,
\beq
m_1 : m_2 : m_3 = \e : \e : 1 \;,
\eeq 
with $m_2^2 \sim m_1^2 \sim \D m_{21}^2 = m_2^2-m_1^2 \sim \e^2$. Since
$\e \sim 0.1$, this pattern is consistent with the LMA solution of the
solar neutrino problem, but not with the LOW solution.

The large $\n_\m$-$\n_\t$ mixing has been obtained as consequence of the 
required very large mass hierarchy (\ref{Mhier}) of the heavy Majorana 
neutrinos. The large $\n_e$-$\n_\m$ mixing follows from the
particular values of parameters ${\cal O}(1)$.
Hence, one expects two large mixing angles, but
single maximal or bi-maximal mixing would require fine tuning. 
On the other hand, one definite prediction is the occurence of exactly 
one small matrix element, $U^{(\n)}_{13} = {\cal O}(\e)$. 
Note, that the obtained pattern of neutrino mixings is independent of the 
off-diagonal elements of the mass matrix $M$. For instance, replacing the 
texture (\ref{Mtext}) by a diagonal matrix, $M = \mbox{diag}(M_1,M_2,M_3)$, 
leads to the same pattern of neutrino mixings.

In order to calculate various observables in neutrino physics we need
the leptonic mixing matrix 
\beq
U = U^{(e)\dg} U^{(\n)}\;,
\eeq
where $U^{(e)}$ is the charged lepton mixing matrix. In our framework we
expect $U^{(e)} \simeq V^{(d)}$, and also $V = V^{(u)\dg} V^{(d)}
\simeq V^{(d)}$ for the CKM matrix since $\e_u < \e_d$. This yields
for the leptonic mixing matrix 
\beq\label{lmix}
U \simeq  V^{\dg} U^{(\n)} \;.
\eeq
To leading order in the Cabibbo angle $\l \simeq 0.2$ we only need the
off-diagonal elements $V^{(d)}_{12} = \Bar{\l} = - V^{(d)*}_{21}$. Since
the matrix $m_d$ is complex, the Cabibbo angle is modified by phases,
$\Bar{\l} = \l\exp{\{i(\phi_d-\a_d)\}}$. The resulting leptonic mixing 
matrix is indeed of the wanted form (\ref{umix}) with all matrix elements 
${\cal O}(1)$, except $U_{13}$,
\beq
U_{13} = \xi s_{23} e^{i(\phi-\b+\j)} - \Bar{\l} s_{23} e^{i(\phi+\b-\j)} 
       =  {\cal O}(\l,\e) \sim 0.1 \;,
\eeq
which is close to the experimental limit. 

Let us now consider the CP violation in neutrino oscillations. Observable 
effects are  controlled by the Jarlskog parameter $J_l$ \cite{jar85}
($\wt{\e}_{ij} = \sum_{k=1}^3 \e_{ijk}$)
\beq
\mbox{Im}\{ U_{\a i}U_{\b j}U_{\a j}^*U_{\b i}^*\} 
= \wt{\e}_{\a\b}\wt{\e}_{ij} J_l\;,
\eeq
for which one finds 
\beq
J_l \simeq  \l s_{12}c_{12}c_{23}s_{23}^2 \sin{(2(\b-\j+\g)+\phi_d-\a_d)}\;.
\eeq
In the case of a small mass difference $\D m_{12}^2$ the CP asymmetry 
$P(\n_\m\rightarrow \n_e) - P(\Bar{\n}_\m\rightarrow \Bar{\n}_e)$
is proportinal to $\d$ (cf.~(\ref{det})). Hence, the dependence of $J_l$ 
on the angle $\g$ is not surprising.

For large mixing, $c_{ij}\simeq s_{ij} \simeq 1/\sqrt{2}$, and
in the special case (\ref{special}) one obtains from the SO(10) phase relation
$\phi-\a = \phi_u -\a_u$ and $\phi_u -\a_u - \phi_d +\a_d = \Delta 
\simeq \pi/2$,
\beq
J_l \simeq {\l\over 4\sqrt{2}} \sin{\left(-{\pi\over 2} + 2\g\right)}\;.
\eeq
For small $\g$ this corresponds to maximal CP violation, but without a
deeper understanding of the fermion mass matrices this case is not
singled out.
Due to the large neutrino mixing angles, $J_l$ is much bigger than the
Jarlskog parameter in the quark sector, $J_q = {\cal O}(\l^6) \sim 10^{-5}$,
which may lead to observable effects at future neutrino factories\cite{blo00}.

According to the seesaw mechanism neutrinos are Majorana fermions. This
can be directly tested in neutrinoless double $\b$-decay. The decay
amplitude is proportional to the complex mass 
\bea
\VEV m &=& \sum_i U_{ei}^2 m_i 
= - (U U^{(\n)\dg} m_\n U^{(\n)*} U^T)_{ee} \simeq 
  - (V^{(d)\dg} m_\n V^{(d)*})_{ee} \NO\\ 
&=& - {1 \over 1 + |\h|^2} \left(\l^2 |\h|^2 e^{2i(\phi_d -\a_d + \b + \phi- \j)}
                             - 2 \l \e e^{i(\phi_d -\a_d + 2 \phi)}\right) m_3\;. 
\eea
With $m_3 \simeq \sqrt{\D m^2_{atm}} \simeq 5\times 10^{-2}$~eV this yields 
$\VEV m \sim  10^{-3}$~eV, more than two orders of magnitude
below the present experimental upper bound\cite{hm99}. 

Finally, consider again the baryon asymmetry which should eventually be related
to the CP violation in neutrino oscillations and quark mixing.
This possibility has recently been discussed also in 
other contexts\cite{jpr01,bmx01}. In the special case\footnote{For the discussion 
of the general case, see ref.\cite{bw01}.} (\ref{special})  one obtains for the 
CP asymmetry,
\beq
\ve_1 \simeq {3\over {16\pi}} \e^6 {|\h|^2\over \s}  
         {(1+|\r|)^2 \over |\h|^2 +|\r|^2}\sin(\phi_u-\a_u) \;.
\eeq  
As expected $\e_1$ depends only on phases of the up-quark matrix and not on
the combination of up$-$ and down$-$quark phases $\D$ which appears in the 
CKM matrix. In addition, the parameter $\s$ enters. Hence, the baryon asymmetry
is not completely determined by properties of the quark matrices and the
CP violation in the neutrino sector.

Numerically, with $\e \sim 0.1$ one has $\ve_1 \sim 10^{-7}$ ,
$|M_1| \simeq (\e^6/|\s|) (1+|\h|^2) v_1^2/m_3 \sim 10^9$~GeV and 
$\wt{m}_1 \sim (|\h|^2+|\r|^2)/(\s(1+|\h|^2)) m_3 \sim 10^{-2}$~eV. 
The baryon asymmetry is then given by
\beq\label{asym}
Y_B \sim  - \k~\mbox{sign}(\s)~\sin{(\phi_u-\a_u)} \times 10^{-9}\;.
\eeq
The parameters $\ve_1$, $M_1$ and $\wt{m}_1$ are rather similar to those 
considered in the previous section. Hence, a solution of the Boltzmann equations
can be expected to yield again a baryon asymmetry in accord with observation.

\section{Conclusions}

Large neutrino mixings, together with the known small quark mixings, have
important implications for the structure of GUTs. In SU(5) models this
difference between the lepton and quark sectors can be explained by lopsided
U(1)$_F$ family symmetries. In these models the heavy Majorana neutrino masses
are not constrained by low energy physics, i.e., light neutrino masses and
mixings. Successful leptogenesis then depends on the choice of the heavy
neutrino masses and is not a generic prediction of the theory.

In SO(10) models the implications of large neutrino mixings are much more
stringent because of the connection between Dirac neutrino and up-quark mass
matrices. It is remarkable that the requirement of large neutrino mixings
determines the relative magnitude of the heavy Majorana neutrino masses in 
terms of the known quark mass hierarchy. This leads to predictions
for neutrino mixings and masses, CP violation in neutrino oscillations and
neutrinoless double $\beta$-decay. The predicted order of magnitude for the
baryon asymmetry is in accord with observation. It would be very interesting
to relate directly the CP violation in the quark sector and in neutrino
oscillations to the baryon asymmetry. This, however, will require a deeper
unterstanding of the quark and lepton mass matrices.

\section{Acknowledgements}

I would like to thank Michael Pl\"umacher, Daniel Wyler and Tsutomu Yanagida 
for an enjoyable collaboration on the topic of this lecture, and I am grateful
to the organisers of the meetings in Dubrovnik and Ustro\'n for their kind hospitality.

\end{document}